\documentclass[fleqn,usenatbib,usedcolumn]{mnras}

\usepackage{newtxtext,newtxmath}
\usepackage[T1]{fontenc}

\DeclareRobustCommand{\VAN}[3]{#2}
\let\VANthebibliography\thebibliography
\def\thebibliography{\DeclareRobustCommand{\VAN}[3]{##3}\VANthebibliography}

\usepackage{graphicx}	% Including figure files
\usepackage{amsmath}	% Advanced maths commands
\usepackage[british]{babel}             % British English hyphenation
\usepackage{newtxtext}                  % Good fonts
\usepackage[T1]{fontenc}                % Good font encoding
\usepackage{amsfonts}
\usepackage{amsbsy}
\usepackage{bm}
\usepackage{url}
\usepackage{microtype}
\usepackage{rotating}
\usepackage{sidecap}
\usepackage{longtable}
\usepackage{lscape}
\usepackage[export]{adjustbox}
\usepackage{xcolor}
\usepackage{colortbl}
\usepackage{booktabs}
\usepackage[flushleft]{threeparttable}

\let\la=\lesssim % for less than similar from newtxmath, not \la from mnras.cls

\def\ergs{{\rm erg\,\,s^{-1}}}
\def\cm2{{\rm cm^{-2}}}

\title[A radio-bright, X-ray obscured GRS~1915+105]{Observations of a radio-bright, X-ray obscured GRS~1915+105} 
\author[S. Motta et al.]{Motta, S.~E.$^{1, 2}$,  Kajava, J.~J.~E.$^{3,4}$, Giustini, M.$^{4}$, Williams, D.~R.~A.$^{2,5}$,  Del Santo, M.$^{6}$, Fender, R.$^{2,7}$,  
\newauthor Green, D.~A.$^{8}$, Heywood, I.$^{2,9,10}$, Rhodes, L.$^{2}$, Segreto, A.$^{6}$, Sivakoff, G.$^{11}$,   Woudt, P.A.$^{7}$\\
$^1$Istituto Nazionale di Astrofisica, Osservatorio Astronomico di Brera, via E.\,Bianchi 46, 23807 Merate (LC), Italy\\
$^2$University of Oxford, Department of Physics, Astrophysics, Denys Wilkinson Building, Keble Road, OX1 3RH, Oxford, United Kingdom\\
$^3$Department of Physics and Astronomy, FI-20014 University of Turku, Finland \\
$^4$Centro de Astrobiolog\'ia (CSIC-INTA), Camino Bajo del Castillo s/n, Villanueva de la Ca\~nada, E-28692 Madrid, Spain \\
$^{5}$ Jodrell Bank Centre for Astrophysics, School of Physics and Astronomy, The University of Manchester, Manchester, M13 9PL, UK\\
$^{6}$ Istituto Nazionale di Astrofisica, IASF Palermo, Via U. La Malfa 153, 90146, Palermo, Italy\\
$^{7}$Department of Astronomy, University of Cape Town, Private Bag X3, Rondebosch 7701, South Africa\\
$^{8}$Astrophysics Group, Cavendish Laboratory, 19 J. J. Thomson Avenue, Cambridge CB3 0HE, UK\\
$^{9}$Department of Physics and Electronics, Rhodes University, PO Box 94, Makhanda 6140, South Africa\\
$^{10}$South African Radio Astronomy Observatory, Cape Town, South Africa. \\
$^{11}$Department of Physics, University of Alberta, CCIS 4-181, Edmonton, AB T6G 2E1, Canada\\
}

\date{Accepted XXX. Received YYY; in original form ZZZ}

\pubyear{2020}

\begin{document}
\label{firstpage}
\pagerange{\pageref{firstpage}--\pageref{lastpage}}
\maketitle

\begin{abstract}
The Galactic black hole transient GRS~1915+105 is famous for its markedly variable X-ray and radio behaviour, and for being the archetypal galactic source of relativistic jets. It entered an X-ray outburst in 1992 and has been active ever since. 
Since 2018 GRS~1915+105 has declined into an extended low-flux X-ray plateau, occasionally interrupted by multi-wavelength flares.  
Here we report the radio and X-ray properties of GRS~1915+105 collected in this new phase, and compare the recent data to historic observations. 
We find that while the X-ray emission remained unprecedentedly low for most of the time following the decline in 2018, the radio emission shows a clear mode change half way through the extended X-ray plateau in 2019 June: from low flux ($\sim$~3mJy) and limited variability, to marked flaring with fluxes two orders of magnitude larger. 
GRS~1915+105 appears to have entered a low-luminosity canonical hard state, and then transitioned to an unusual accretion phase, characterised by heavy X-ray absorption/obscuration. Hence, we argue that a local absorber hides from the observer the accretion processes feeding the variable jet responsible for the radio flaring. The radio--X-ray correlation suggests that the current low X-ray flux state may be a signature of a super-Eddington state akin to the X-ray binaries SS433 or V404 Cyg.
\end{abstract}


%%%%%%%%%%%%%%%%%%%%%%%%%%%%%%%%%%%%%%%%%%%%%%%%%%

%%%%%%%%%%%%%%%%% BODY OF PAPER %%%%%%%%%%%%%%%%%%

%%%%%%%%%%%%%%%%%%%%%%%%%%%%%%%%%%%%%%%%%%%%%%%%%%%%%%%%%%%%%%%%%%%%%%%%%%%%%%%%%%%%%%%%%%%%%%%%%%%%%%%%%%%%%%%%%%%%%%%%%%%%%%%%%%%%%%%
\section{Introduction}

In black hole (BH) X-ray binaries a stellar mass black hole accretes via an accretion disc formed by matter stripped from a low-mass companion star. BH X-ray binaries are typically transient systems, i.e. they alternate between long states of quiescence, characterised by a luminosity typically of the order of $L \sim 10^{34}\,\ergs$  (see \citealt{Wijnands2015}), and relatively short outbursts, during which their luminosity can reach $\sim 10^{39}\,\ergs$. During outbursts these systems show clear repeating patterns of behaviour across various accretion states, each associated with mechanical feedback in the form of winds and relativistic jets (e.g., \citealt{Fender2009}, \citealt{Ponti2012}).

The hard states, characterised by highly variable X-ray emission dominated by hard photons, are associated with steady radio jets \citep{Fender2004}. In a few occasions, cold (i.e., consistent with being not ionised) winds, which appear to co-exist with the radio jets, have been observed in the optical band during the hard state \citep{Munoz-Darias2016}, casting doubts on the idea according to which jets and winds are associated to different accretion states, and therefore cannot co-exist.
In the X-ray low-variability soft states, X-ray spectra are dominated by thermal emission from a geometrically thin, optically thick accretion disc, the radio emission is quenched\footnote{Any residual radio emission in the soft state has been so far associated with ejecta launched before the transition to the soft state (see e.g. \citealt{Bright2020}). }, and X-ray (ionised) winds are seen \citep{Ponti2014,Tetarenko2018}. In between these two states lie the intermediate states, with properties in between the hard and the soft state, and during which short-lived, powerful relativistic radio ejections are observed \citep{Fender2009}.
Quasi-simultaneous X-ray and radio observations of BH X-ray binaries have been fundamental to the study of the connection between the accretion and the jet production mechanism, which led to the establishment of a disc-jet coupling paradigm \citep{Fender2004}. Such a coupling gives rise in the hard state to a well-known non-linear relation between the X-ray and the radio luminosity, known as the radio--X-ray correlation (e.g., \citealt{Gallo2003} and \citealt{Corbel2003}), which also encompasses AGN when a mass scaling term is considered \citep{Merloni2003,Falcke2004, Plotkin2012, Gultekin2019}. 

%\subsection{GRS~1915+105}
\smallskip

GRS~1915+105 is one of the most well studied Galactic BH X-ray binaries, which firsts appeared as a bright transient in August 1992 and remained very bright in X-rays and in radio until recently \citep{Negoro2018}. GRS 1915+05  was the first Galactic source observed to display relativistic super-luminal radio ejections \cite{Mirabel1994}, and it is still considered the archetypal galactic source of relativistic jets. 
This system, located at a radio parallax distance of 8.6$^{+2.0}_{-1.6}$ kpc, hosts a stellar mass black hole (12.4$^{+2.0}_{-1.8}$M$_{\odot}$  \citealt{Reid2014}), believed to accrete erratically close to the Eddington limit.  Several characteristic X-ray variability patterns observed in the X-ray light curve of GRS~1915+105 \citep{Belloni2000} are believed to reflect transitions from and to three accretion states: two soft states (A and B), and a hard state, C, all slightly different from the \textit{canonical} states seen in other BH binaries \citep{Belloni2016}. State A and B are characterised by limited X-ray variability, and a substantial contribution from an accretion disc with a variable temperature that can reach 2~keV. State C shows high X-ray variability and no disc contribution to the X-ray spectrum, and is known to be associated with steady radio jets \citep{Rushton2010}. Such jets appear as flat-top periods in the radio light curve, characterised by relatively high radio flux densities ($\sim$~100~mJy beam$^{-1}$), an optically thick radio spectrum, and a flat low-flux X-ray light curve (\citealt{Pooley1997}). In the radio--X-ray plane, state C corresponds to a high-luminosity extension of the radio--X-ray correlation \citep{Gallo2003}.
Radio Plateaus are generally preceded and followed by flaring periods due to the launch of relativistic ejections (\citealt{Rodriguez1999}), which have been repeatedly resolved as extended radio jets on a range of scales from $\sim$1 mas to hundreds of arcseconds \citep{Dhawan2000,Miller-Jones2005,Rushton2007,Fender1999,Miller-Jones2007}.

While GRS~1915+105 is in many ways unique, it shares many properties with more conventional transient black-hole binaries (such as GX 339--4, see, e.g., \citealt{Done2004}, \citealt{Soleri2008}), and also with some quasars \citep{Marscher2002,Chatterjee2009}, which display mass-scaled versions of GRS~1915+105's correlated X-ray and radio behaviour close to jet ejection events. Hence, comprehending the coupling between inflow and outflow in GRS~1915+105 is important not only for X-ray binary systems, but has broader relevance
for studies of active galactic nuclei, and potentially of all the jetted systems powered by accreting compact objects \citep{Fender2004}.

In 2018 July ($\sim$MJD 58300), after 26 years of extreme 
X-ray and radio activity, GRS~1915+105 entered an unusually long period of low-flux in the X-rays and radio  \citep{Negoro2018,Motta2019}, which made some to believe that quiescence was close. 
Around the end of  2019 March (MJD 58600), GRS~1915+105 entered a sudden further X-ray dimming \citep{Homan2019,Rodriguez2019}, which reinforced the hypothesis that the 26-year long outburst of GRS 1915+105 was nearing an end. However, only days later, on 2019 May 14th (MJD 58617), renewed flaring activity at different wavelengths appeared to invalidate the quiescence hypothesis. After approximately a month of marked multi-wavelength activity \citep{Iwakiri2019, Miller2019a, Neilsen2019, Jithesh2019, Vishal2019, Svinkin2019, Koljonen2019, Balakrishnan2019, Trushkin2019,Motta2019}, GRS~1915+105 entered a new X-ray low-flux state. % \SEM{Yes, these are all needed, I didn't even put them all, I selected the most relevant ones}
X-ray observations with Swift, NuStar and Chandra taken during this second low phase showed hard spectra characterised by heavy and occasionally partial covering absorption with equivalent column densities  $N_{\rm H} $>$ 3 \times 10^{23}$ cm$^{-2}$, i.e. over an order of magnitude larger then the usual equivalent column density in the direction of GRS 1915+105 \citep{Miller2019,Koljonen2020,Miller2020, Balakrishnan2021}. 
The presence of intrinsic absorption in GRS~1915+105 has never previously been reported, and have been observed only rarely in other X-ray binaries. Two notable exceptions are the BH V404 Cyg (see, e.g., \citealt{Zycki1999} and \citealt{Motta2017a}), which showed clear signatures of heavy and variable intrinsic absorption during both the outbursts monitored in the X-rays, and SS 433, which is believed to be obscured by its own inflated accretion disc \citep{Fabrika2004}. In both these systems obscuration was the consequence of erratic (V404 Cyg) or sustained (SS 433) super-Eddington accretion, which in both cases is associated with extreme activity of the radio jets \citep{Spencer1979,Miller-Jones2019}.

In this paper we report on the behaviour of GRS~1915+105 based on the long-term monitoring operated by a number of X-ray All-sky monitors and radio facilities. We compare the recent (as in 2020) evolution of the systems with its past behaviour, with the aim of highlighting the peculiarities of the current, highly unusual state. In Section \ref{Sec:Obs} we describe our data reduction and analysis, in Sec. \ref{sec:results} we present our results, and in Sec. \ref{sec:discussion} we will discuss our findings. Finally, in Sec. \ref{sec:conclusions} we will summarise our main results and outline our conclusions. 

%%%%%%%%%%%%%%%%%%%%%%%%%%%%%%%%%%%%%%%%%%%%%%%%%%%%%%%%%%%%%%%%%%%%%%%%%%%%%%%%%%%%%%%%%%%%%%%%%%%%%%%%%%%%%%%%%%%%%%%%%%%%%%%%%%%%%%%
\section{Observations and data Analysis}\label{Sec:Obs}

In this section we describe the reduction and analysis of the data from the radio and X-ray facilities used in this work. A log of the data considered is given in Tab. \ref{tab:log}.

%----------------------
\begin{table}
 \centering
\begin{tabular}{c c c }
\hline \hline
Instrument		& Energy/Frequency 			& Time covered 	(MJD)			\\ 
\hline \hline
Ryle Telescope	   &  15.5~GHz (350 MHz)     		&   49856--53898		\\
AMI-LA  		   &  15.5~GHz (5~GHz)     		&   54615--56925		\\
MeerKAT 		   &  1.28 GHz (0.86~GHz)    	&   57642--58926     	\\
\hline
RXTE/ASM           &  2--12 keV   	&   50088--55859     	\\
MAXI/GSC           &  2--12 keV  	&   55054--59169     	\\
\textit{Swift}/BAT &  15--50 keV   	&   53347--59169     	\\

\hline
\end{tabular}
\caption{A log of the data used in this work. For MeerKAT, the Ryle telescope and AMI-LA we give in parenthesis the bandwidth used. }\label{tab:log}
\end{table}
%----------------------

\subsection{Radio}
\subsubsection{Ryle and AMI-LA telescopes}\label{sec:AMI}

From 1995 May to 2006 June (MJD 49850 to 53900) the Ryle telescope routinely observed GRS~1915+105 as part of an extensive monitoring campaign on a number of bright radio transients. 
In 2006 the Ryle telescope was partly converted into the Arcminute Microkelvin Imager Large Array (AMI-LA) and observations of GRS~1915+105 continued until 2016 January (MJD 57400), when the array was switched off to allow the original analog correlator to be upgraded with a digital one.
Observations resumed in 2016 June (MJD 57640) and continued until March 2020, when AMI-LA had to be shut down due to the Covid-19 outbreak (MJD 58926). 

Data from the Ryle telecope have been published by, e.g., \citet{Pooley1997}, \citet{Klein-Wolt2002}, and \citet{Rushton2010}, and we refer the reader to those works for details on the data reduction. 
The AMI-LA observations were conducted at a central frequency of 15.5 GHz with a 5 GHz bandwidth, divided into 8 channels for
imaging (for the digital correlator data there were originally 4096 narrow channels).  We used 3C286 as the flux/bandpass calibrator, and J1922+1530 as the interleaved phase calibrator.
We reduced the data with a custom pipeline that uses the AMI \textsc{reduce\_dc} software, which automatically flags for radio frequency interference (RFI), antenna shadowing, and hardware errors, performs Fourier transforms of the lag-delay data into frequency channels (for the analog
correlator data), and then applies phase and amplitude calibrations (e.g., \citealt{Perrott2013}). 
We carried out further flagging using the Common Astronomical Software Applications (\textsc{CASA}) package (\citealt{McMullin2007}), which was also used for the interactive cleaning. 
For imaging we use natural weighting with a clean gain of 0.1. To measure the source flux density we use the CASA task \textsc{imfit}. The synthesised beam of the AMI-LA when observing at the declination of GRS 1915+105 is 40 arcsec $\times$30 arcsec. The target is
unresolved in all epochs.

The data have been binned differently based on the brightness of the target. When the target was relatively faint (flux density $<$10~mJy beam$^{-1}$) we report the average flux measured in each epoch, which have a variable total duration of 1 to 7 hr. When the target was brighter, we split each epoch into shorter segments (down to 6-minutes long), depending on the source flux.

\subsubsection{MeerKAT telescope}

As part of the ThunderKAT large survey project \citep{Fender2016b} we observed  GRS~1915+105 with the MeerKAT radio interferometer 38 times. Data were obtained at a central frequency of 1.28 GHz across a 0.86 GHz bandwidth consisting of either 4096 channels or 32768 channels (in this second case data were binned for consistency to 4096 channels before any further analysis). 
Observations covered the period between 2018 December and 2020 November  (MJD 58460 -- 59168). 
We initially observed the target every several weeks. When AMI-LA stopped operations, we switched to a weekly monitoring of GRS~1915+105 with MeerKAT.  
The first MeerKAT observation consisted of an observation with a total duration 90 min, of which 60 min is on-source, 20 minutes is on the flux and band-pass calibrator, and 3 minutes on the phase calibrator. 
All other observations had a total on-source integration time of 15 minutes, and a flux and bandpass calibrator, and phase calibrator times of 10 and 4 minutes, respectively. 
We used J1939$-$6342 as the flux and band-pass calibrator, and J2011$-$0644 as the complex gain calibrator. Between 58 and 63 of the 64 available dishes were used in the observations, with a maximum baseline of 7.698 km. 

The subsequent analysis has been conducted via a set of \textsc{Python} scripts specifically tailored for the semi-automatic processing of MeerKAT data (\textsc{OxKAT}\footnote{\url{https://github.com/IanHeywood/oxkat}}, \citealt{Heywood2020}). We used \textsc{CASA} to flag the first and final 100 channels from the observing band, autocorrelations and zero amplitude visibilities. Then we further flagged  the data to remove RFI in the time and frequency domain. Flux density scaling, bandpass calibration and complex gain calibration were all performed within \textsc{CASA} using standard procedures. A spectral model for the phase calibrator is derived starting from the flux and band-pass calibrator, by temporarily binning the data into 8 equal spectral windows. 
We then averaged the data in time (8~s) and frequency (8 channels) for imaging purposes, and we used \textsc{WSClean} (\citealt{Offringa2012}) to image the entire MeerKAT square degree field. 
%We used Briggs weighing and a robust parameter of $-0.3$ and a clean gain of 0.1. In order to efficiently clean the crowded field of GRS~1915+105 we applied a mask to minimise the artefacts due to the presence of bright, extended sources in the field. In order to improve the quality of our images, we performed self-calibration of the phases within \textsc{CASA}.

We measured the fluxes averaging data in each epoch, so that each MeerKAT point (blue diamonds in Fig. \ref{fig:MAXI_BAT}, panel \textit{(a)}) corresponds to a 15 minutes of on-target time, except the first point (MJD 58460), which corresponds to a 60-min integration time. The target is unresolved in all observations, and in this paper we will only consider the flux densities measured in the MeerKAT images using the \textsc{imfit} task in \textsc{CASA}. More in-depth analysis of the MeerKAT maps will be presented in a dedicated paper (Motta et al. in prep).

\subsection{X-ray All-Sky monitors}

We extracted long-term light curves for GRS~1915+105 using the public data available on the web pages of RXTE/ASM (ASM\footnote{\url{http://xte.mit.edu/ASM\_lc.html}}) and MAXI/GSC (MAXI\footnote{\url{http://maxi.riken.jp/top/lc.html}}), and from the survey data collected with the BAT telescope on board the Neil Gehrels Swift Observatory (\textit{Swift}). 
We used the MAXI on-demand tool\footnote{\url{http://maxi.riken.jp/mxondem/}} to extract the data covering the 2--12 keV band in order to be able to directly compare the light curves from MAXI and the ASM. We converted the ASM and MAXI count rates into fluxes using an approximate counts-to-flux conversion fraction, based on the mean count rates of the Crab, which corresponds to 75 count~s$^{-1}$for the ASM and 3.74 count~s$^{-1}$for the MAXI, respectively. We note that such a count rate to flux conversion is not rigorous, but it is sufficient for our purposes. We account for any bias introduced by the conversion assuming a conservative uncertainty on the flux of 20 per cent.
Owing to the larger energy interval covered by BAT (nominally 15--150 keV), a count rate to flux conversion would not be accurate when using the Swift/BAT transient monitor results \citep{Krimm2013}. Hence, we processed the BAT survey data retrieved from the HEASARC public archive by using the \textsc{BatImager} code developed by \cite{Segreto2010}, which is dedicated to the processing of coded mask instrument data.  \textsc{BatImager} performs image reconstruction via cross-correlation and, for each detected source, generates light curves and spectra. 
We processed BAT survey data from MJD 53347 to MJD 59169, and extracted one spectrum per day using the official BAT spectral redistribution matrix and a logarithmic binned energy grid. Then, we fit the spectra from 15 keV to 50 keV with a simple power-law and derived the observed flux. 
Using the fluxes obtained as described, we calculated a X-ray colour $C$ = $F_{\rm hard}/F_{\rm soft}$, where $F_{\rm hard}$ is the flux coming from BAT, and $F_{\rm soft}$ is the flux coming from either the ASM or MAXI. Higher colour means a harder spectrum.
The ASM and MAXI data overlap with those from BAT data by several years, and overlap to each other for about two years (i.e. from MJD 55054 to 55859). This allows us to confirm that ASM and MAXI provides consistent data (see \ref{fig:MAXI_BAT}, panel \textit{(b)}). 
Variability on time-scales significantly shorter than a day is known to occur during various accretion states both in the X-rays \citep{Belloni2000} and in radio \citep{Pooley1997} in the flux from GRS~1915+105. However, the aim of this work is to study the long term behaviour of this system. Thus we focused on the variability occurring on time-scales longer than a few days, rather than the details of a particular flare. Therefore, we rebinned the ASM, MAXI and BAT light curves to the same 1-day long time bins. 

We also extracted energy spectra in specific time intervals (see Sec. \ref{sec:spectra}) using the on-demand MAXI tool with the default extraction parameters, and specifying the good time intervals to be used for the extraction. The spectra were then fitted within \textsc{xspec} (v 12.11.0).

\begin{figure*}
\centering
\includegraphics[width=0.99\textwidth]{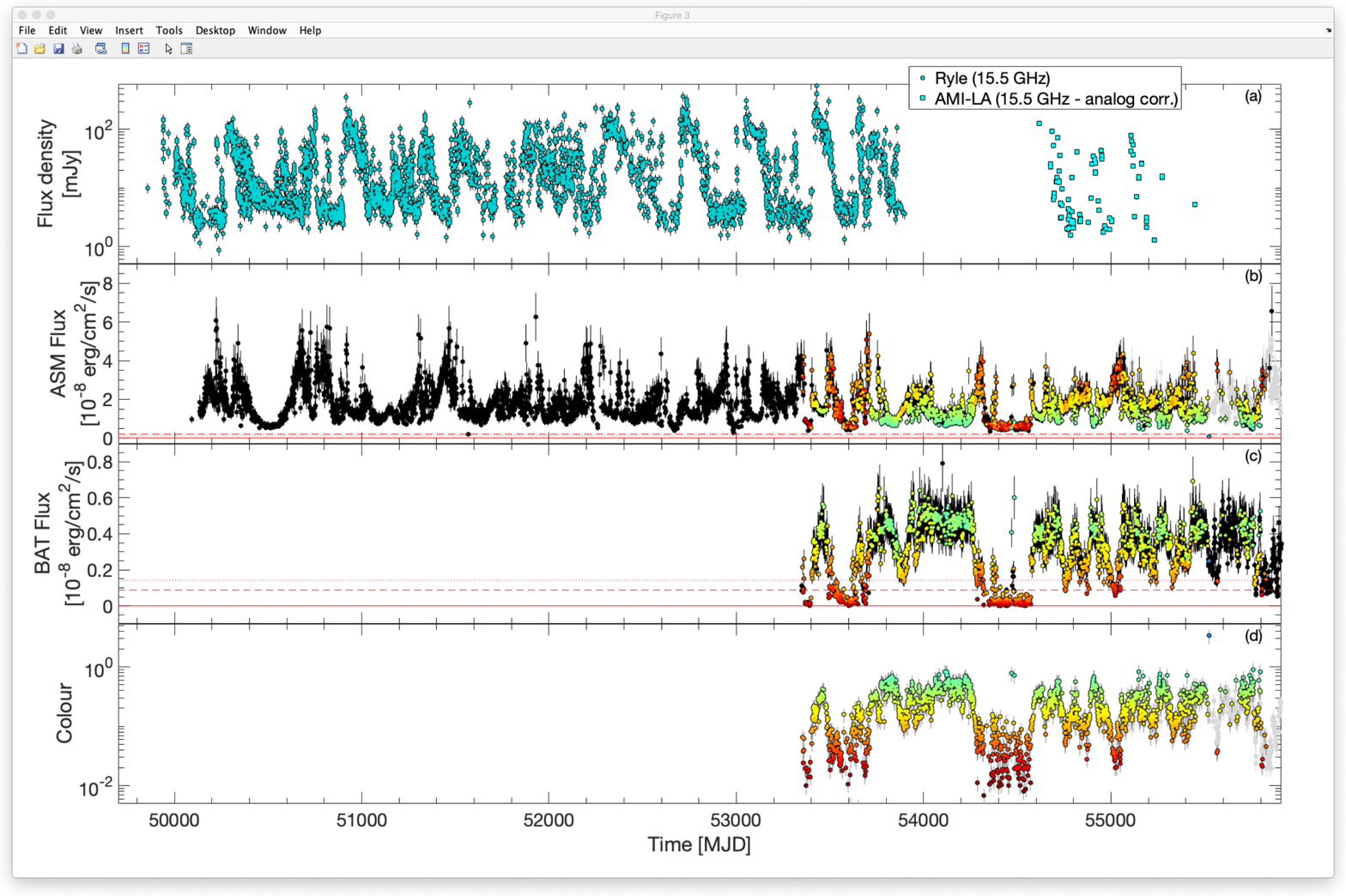}
\caption{AMI-LA and Ryle data (panel \textit{a}), RXTE/ASM data (panel \textit{b}) and BAT data (panel \textit{c}), covering over 16 years. Panel \textit{b}, \textit{c} and \textit{d} are colour-coded based on the spectral X-ray colour displayed panel \textit{d}, calculated as ratio between the BAT and the ASM fluxes. Redder points correspond to softer spectra.
The grey points in panel \textit{b} are from MAXI (the same as in Fig. \ref{fig:MAXI_BAT}, panel \textit{b}) and are plotted to allow a comparison with the ASM data. Similarly, the grey points in panel \textit{d} correspond to the colours shown in Fig. \ref{fig:MAXI_BAT} for comparison. }\label{fig:ASM_BAT}
\end{figure*}

\begin{figure*}
\centering
\includegraphics[width=0.99\textwidth]{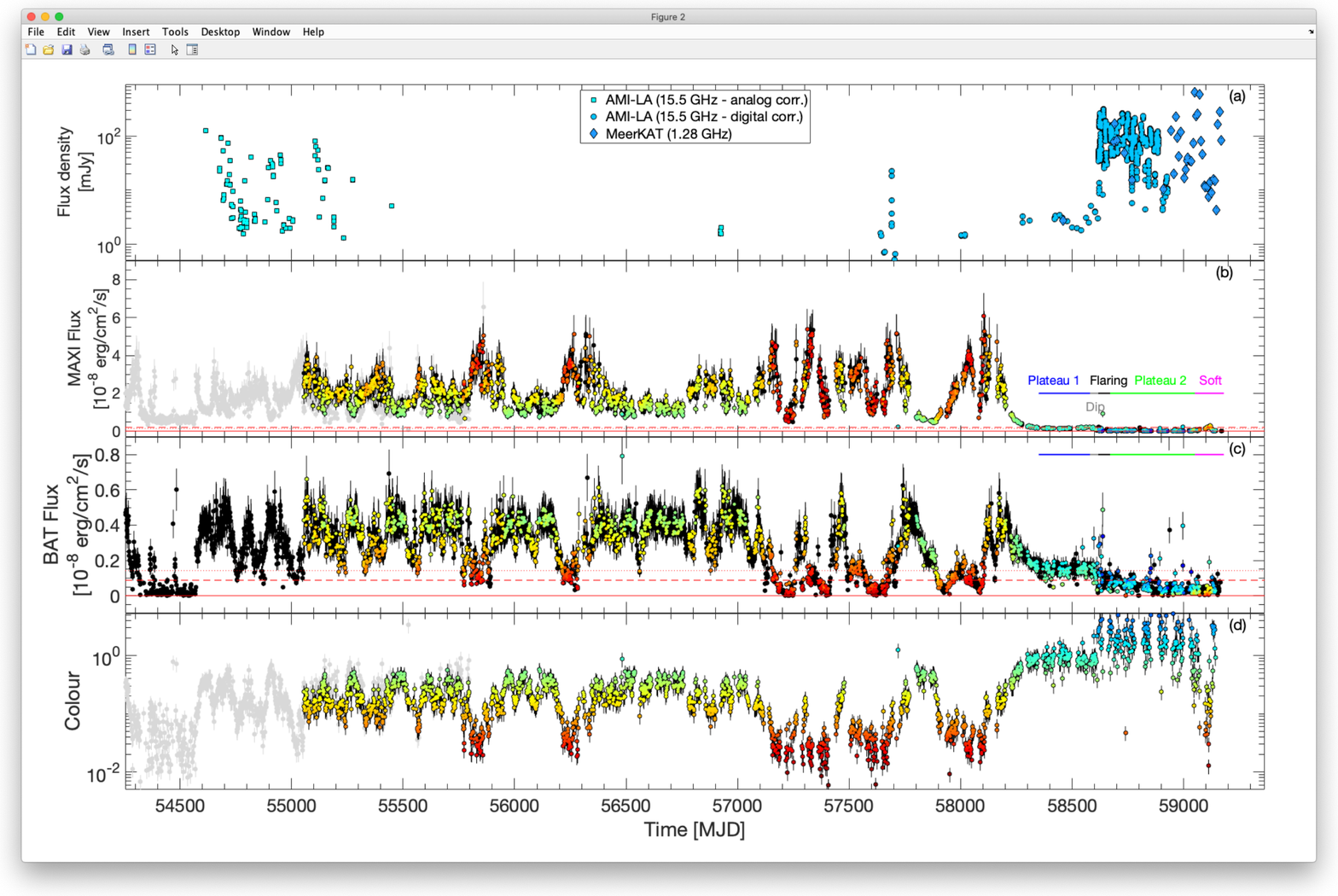}
\caption{AMI-LA and MeerKAT data (panel \textit{a}), MAXI data (panel \textit{b}), and BAT data (panel \textit{c}), covering approximately 14 years. The  colour-coding is the same as in Fig. \ref{fig:ASM_BAT}, with the difference that the spectral X-ray colour is calculated as the ratio between the BAT and the MAXI fluxes. The grey points in panel \textit{b} are from the ASM (the same as in Fig. \ref{fig:ASM_BAT}, panel \textit{b}) and are plotted to allow a comparison with the MAXI data. Similarly, the grey points in panel \textit{d} correspond to the colours shown in Fig. \ref{fig:ASM_BAT} for comparison. }\label{fig:MAXI_BAT}
\end{figure*}

\begin{figure}
%\begin{figure*}
\centering
\includegraphics[width=0.48\textwidth]{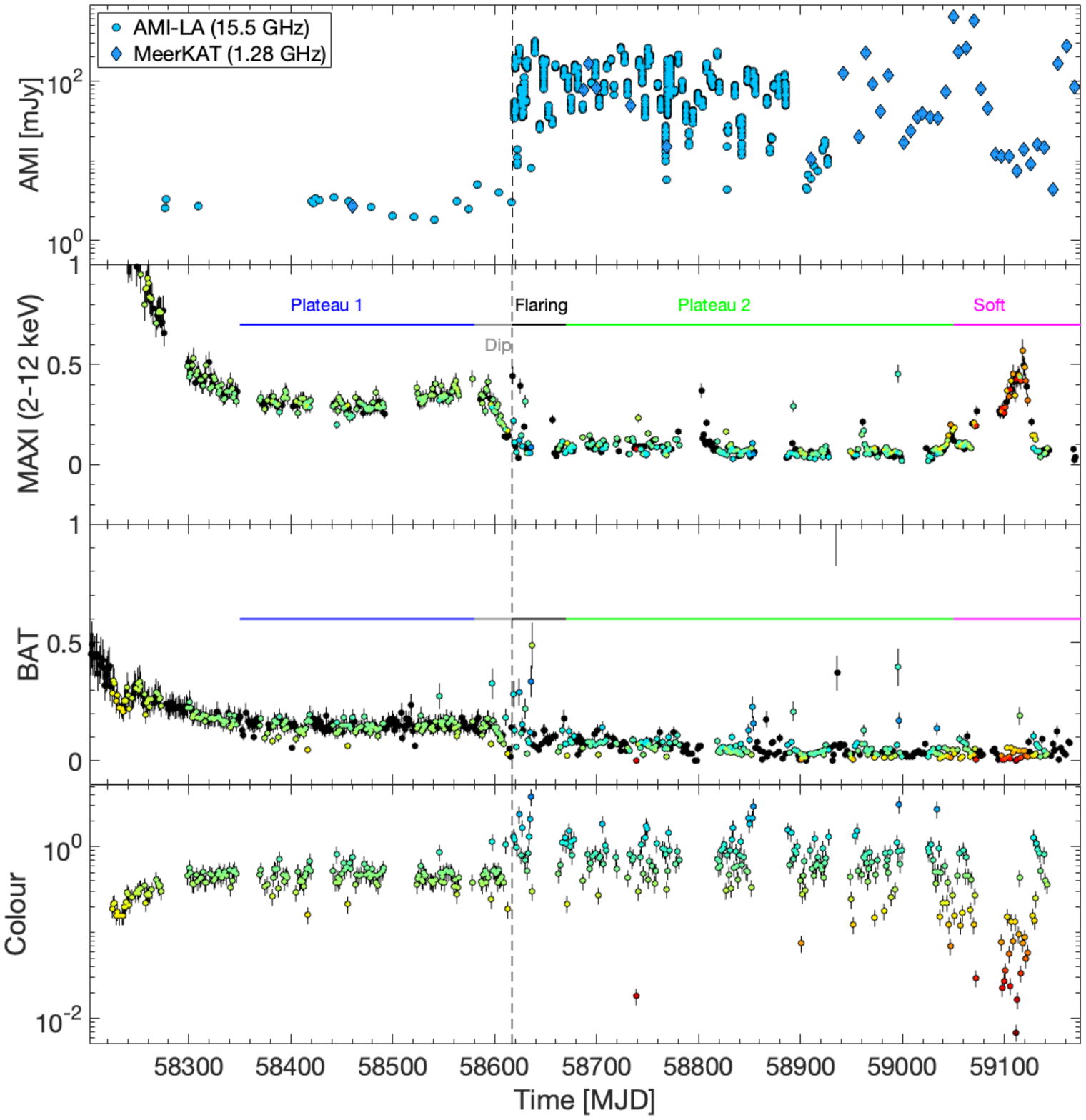}
\caption{A zoom of Fig. \ref{fig:MAXI_BAT}, focusing on the most recent evolution of GRS~1915+105. The vertical dashed line marks the time of the change in the radio behaviour occurred on $\approx$ MJD 58617.}\label{fig:MAXI_BAT_zoom}
%\end{figure*}
\end{figure}

\begin{figure}
\centering
\includegraphics[width=0.48\textwidth]{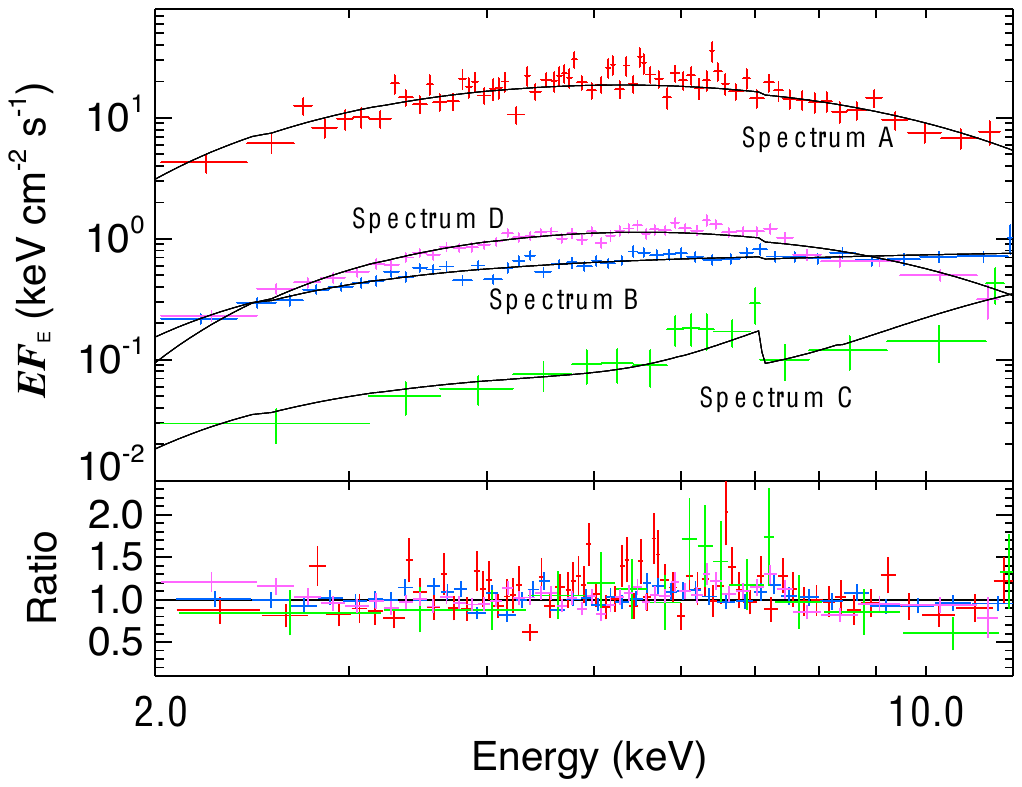}
\caption{MAXI unfolded average spectra extracted during different phases of the evolution of GRS~1915+105, and the ratios to the best fits. Spectra are taken around the peak of the soft flare preceding Plateau 1 (Spectrum A, red); during Plateau 1 (Spectrum B, blue), during Plateau 2 in a time interval flare-free (Spectrum C, green), during the Soft Phase (Spectrum D, magenta). The best-fit parameters are listed in Tab. \ref{tab:para}. }\label{fig:spectra}
\end{figure}

\begin{figure}
\centering 
\includegraphics[width=0.45\textwidth]{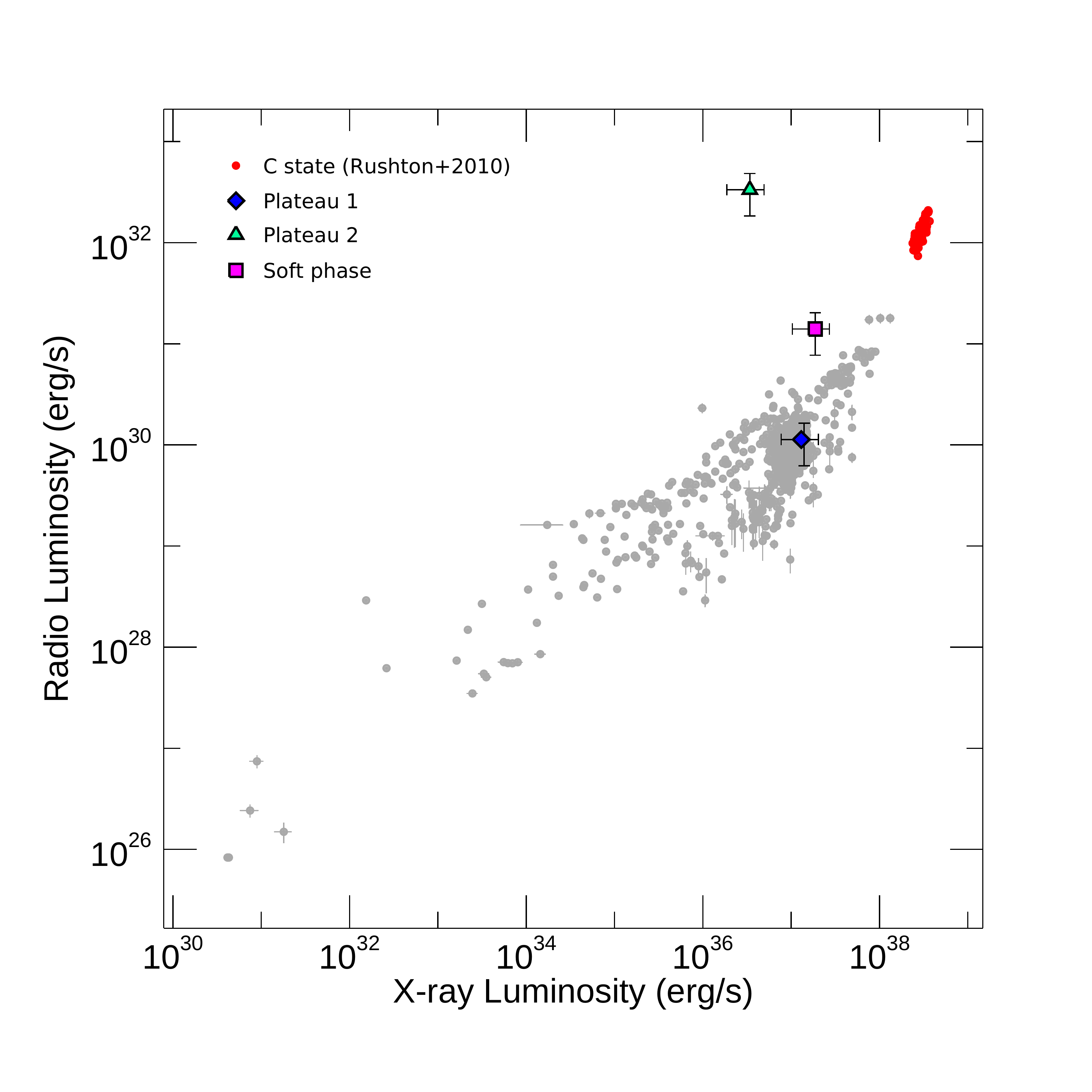}
\caption{The radio--X-ray plane. We mark in grey the fluxes from all the BH transients considered by Motta et al. (2017a), plus MAXI J1820+070, based on Bright et al. (2020). The red dots mark the data from Rushton et al. (2010), who selected data corresponding to a a bright hard state, or state C. The three symbols mark the three main phases described in Sec. \ref{sec:results}: Plateau 1, Plateau 2, and soft.
}\label{fig:RX_corr}
\end{figure}

%%%%%%%%%%%%%%%%%%%%%%%%%%%%%%%%%%%%%%%%%%%%%%%%%%%%%%%%%%%%%%%%%%%%%%%%%%%%%%%%%%%%%%%%%%%%%%%%%%%%%%%%%%%%%%%%%%%%%%%%%%%%%%%%%%%%%%%
\section{Results}\label{sec:results}

\subsection{Overall behaviour}

Figure \ref{fig:ASM_BAT} displays, from the top: the radio light curve taken at a central frequency of 15.5 GHz (Ryle Telescope and AMI-LA data, light and dark cyan points), the soft X-ray light curve (ASM data covering the 2--12 keV band), the hard X-ray light curve (BAT data covering the 15-50 keV band), and the X-ray colour. Figure \ref{fig:MAXI_BAT} shows, from the top: the radio light-curve from data taken at a central frequency of 15.5 GHz (AMI-LA, clear blue points) and 1.28 GHz (MeerKAT, blue diamonds), the soft X-ray light curve (MAXI, covering the 2--12 keV band), and again the BAT light curve and the X-ray colour.
Note that the two figures are plotted using similar time scales, and overlap by several years, but were kept separated to allow for the inspection of both the ASM and MAXI data.
In both figures, all panels except the top ones are colour-coded so that redder corresponds to a softer spectrum (displayed in panel (\textit{d}) in both figures). Wherever the ASM or MAXI data did not overlap with the BAT data, we left the points black.  Part of the radio and X-ray data presented in Fig. \ref{fig:ASM_BAT} have been already published by, e.g.,  \citet{Pooley1997},  \citet{Fender1999}, \citet{Pooley2010}, \citet{Klein-Wolt2002} and \citet{Rushton2010}. % For more details of the past radio and X-ray evolution of GRS~1915+105 we refer the reader to the above works, and to \citet{Fender2004a} for an extensive review. % on GRS~1915+105.

All the light curves are characterised by periods of intense flaring, interleaved by relatively short and quiet phases, both in the X-rays and in radio. 
In Fig.  \ref{fig:MAXI_BAT} we can easily identify a time when both the radio and the X-ray behaviour of GRS~1915+105 changed, i.e.  around MJD 58300 (2018 July), when the source entered a first low-flux phase approximately 11 months-long, to which we refer to as \textit{Plateau 1}. The average flux level observed during Plateau 1 is approximately $F \approx 0.30 \times 10^{-8}\, {\rm erg\,cm^{-2}\,s^{-1}}$ and $F \approx 0.15 \times 10^{-8}\, {\rm erg\,cm^{-2}\,s^{-1}}$ in the MAXI and the BAT data, respectively, and is marked with a dotted line in panels \textit{(b)} and \textit{(c)} in both Fig. \ref{fig:ASM_BAT} and Fig. \ref{fig:MAXI_BAT} for comparison. According to the X-ray colour plotted in  panel (\textit{d}) of Fig. \ref{fig:MAXI_BAT}, this decay led the source from a relatively soft state (around MJD 58000, with colour of $\approx$0.05) to a significantly harder state, characterised by an X-ray colour of $\approx$1 (cyan in Fig. \ref{fig:MAXI_BAT} and \ref{fig:MAXI_BAT_zoom}). 
The AMI-LA data show that in the radio band Plateau 1 corresponds to relatively low flux densities between 1 and 5~mJy beam$^{-1}$, consistent with the lower end of the radio flux densities observed over the 11 years of activity covered by the Ryle telescope.

On MJD 58613 GRS~1915+105 showed a fast decay to an even lower X-ray flux -- we refer to this phase as \textit{Pre-Flare Dip}. The Pre-Flare Dip can be easily discerned both in the MAXI and in the BAT data, as shown in Fig. \ref{fig:MAXI_BAT_zoom}. Shortly afterwards, GRS 1915+105  entered a multi-band flaring period -- which we refer to as the \textit{Flaring Phase} -- that in the X-rays lasted approximately 1 month.
The Flaring Phase, instead, appears as a few isolated points in both the MAXI and BAT light curves around MJD 58600. 
The inspection of the orbit-by-orbit BAT light curve (not displayed\footnote{The orbit-by-orbit BAT light curve is available at \url{http://swift.gsfc.nasa.gov/results/transients/GRS1915p105/}}) shows that the flares in this phase have a variable duration of a few hours up to a few days. 
Given the coarse 1 day binning that we employed to compare the MAXI and BAT data, the X-ray  colour displayed in Fig. \ref{fig:MAXI_BAT} is not sensitive to the fast spectral changes occurring during this high-variability phase. Thus, the colour does not provide any specific information on the spectral properties of the flares, apart from the fact that they appear to be predominantly hard, hence more pronounced in the BAT curve.

GRS~1915+105 subsequently entered a new X-ray plateau (\textit{Plateau 2}) around MJD 58700, which was  occasionally interrupted by short flares lasting approximately 1 day or less (see also \citealt{Neilsen2020}), again visible as isolated points in the BAT and MAXI light curves. This second plateau lasted over 13 months. 
Interestingly, the MeerKAT and AMI data show that the radio flaring did not cease with the X-ray and multi-band flaring, but continued for several months, until at least MJD 59100.
In contrast, the X-ray flux level continued to slowly decline from $F\approx0.08\times 10^{-8}\, {\rm erg\,cm^{-2}\,s^{-1}}$ to $F\approx0.06\times 10^{-8} \, {\rm erg\,cm^{-2}\,s^{-1}}$ in the MAXI data (red dashed and red solid line in panel \textit{(b)} in both Fig. \ref{fig:ASM_BAT} and Fig. \ref{fig:MAXI_BAT}), and from $F\approx0.08\times 10^{-8}\, {\rm erg\,cm^{-2}\,s^{-1}}$ to $F\approx0.03\times 10^{-8}\, {\rm erg\,cm^{-2}\,s^{-1}}$ in the BAT data (red dashed and red solid line in panel \textit{(c)} in the same figures). The signal-to-noise ratio of both the data from MAXI and BAT was very limited in this phase, due to the low count rates from the source, but the X-ray colours measured in this phase suggest a markedly hard spectral shape. 
The radio flaring sampled by AMI-LA in this second plateau does not qualitatively differ from that observed previously over almost three decades\footnote{Noted that MeerKAT observes at a lower frequency than AMI-LA, and a optically thin to optically thick flare would peak sooner and higher at 15.5~GHz than at 1.28~GHz \citep{vanderlaan1966}.}.  The radio emission is characterised by flares of variable amplitude and duration, spanning a flux range between 3 and 300~mJy beam$^{-1}$ in the AMI-LA data, and up to 900~mJy beam$^{-1}$ in the MeerKAT data, and variability on time scales from minutes to several hours. 

More recently, around $\sim$ MJD 59050, the MAXI light curve and the evolution of the colour in Fig. \ref{fig:MAXI_BAT} and Fig. \ref{fig:MAXI_BAT_zoom} shows that GRS~1915+105 returned to a flux comparable to that of Plateau 1, but with a much softer spectrum characterised by an X-ray colour of approximately 0.05. This indicates that GRS 19105+105 has likely transitioned to a significantly softer state, which, coherently with what was observed in the past (see, e.g., data around MJD 53600 in Fig. \ref{fig:ASM_BAT}), features a  diminished radio activity, with flux densities of approximately 10~mJy beam$^{-1}$ and limited variability. 
This last Soft Phase lasted until $\approx$MJD 59140, when both the soft and hard X-ray flux dropped to values comparable to those observed during Plateau 2, and the radio flaring resumed, qualitatively similar to what observed prior to the softening. 

%--- to results
At no point in the past has GRS~1915+105 reached fluxes as low as those measured during Plateau 1 and 2 (see also \citealt{Negoro2018}), despite the presence of several low-flux phases observed both in the soft and in the hard X-rays, all in general associated with relatively low radio flux and low colours, indicative of a relatively soft state (see also \citealt{Klein-Wolt2002} and \citealt{Fender2004a}). The soft plateau occurred around MJD 54500 is, however, noteworthy. The lack of radio observations during such a plateau unfortunately prevents us from drawing solid conclusions on the state GRS~1915+105 was in, but it is possible that the source entered a relatively soft state (as indicated by the steep photon index measured) similar to that sampled by the source around MJD 59000.

\subsection{Spectral analysis}\label{sec:spectra}

To further investigate the properties of the emission from GRS~1915+105 over the phases we described, we extracted four time-averaged MAXI energy spectra in specific phases of the evolution of the source, with variable exposure times chosen to avoid times of variable emission, and to guarantee similar signal-to-noise ratios. Figure \ref{fig:spectra} shows the spectra we obtained and their best fit: 
Spectrum A (in red, extracted in the time interval between from MJD 58034 and 58035), extracted at the peak of a soft flare preceding Plateau 1;
Spectrum B, extracted during Plateau 1 (in blue,  MJD 58450--58490); 
Spectrum C, extracted during Plateau 2 in a time interval flare-free (in green, MJD 58830--58850); 
Spectrum D, extracted during the Soft Phase (in magenta, MJD 59095--59116).
The the best fit parameters are reported in Tab. \ref{tab:para}. 

We fitted the four spectra with phenomenological models constituted by either a powerlaw, or a disc blackbody continuum, each modified by interstellar absorption (\textsc{tbfeo} in \textsc{xspec}), and an additional partially covering absorber (\textsc{tbpcf} in \textsc{xspec}, see \citealt{Wilms2000}), so that the models used in \textsc{xspec} have the form: \texttt{tbfeo $\times$tbpcf$\times$(powerlaw)} or \texttt{tbfeo$\times$ tbpcf$\times$(discbb)}.
We fixed the interstellar absorption parameter to $N_{\rm H} = 5.0\times10^{22}$ cm$^{-2}$ \citep{Miller2016,Zoghbi2016}.
In order to compare the two soft spectra (A and D) and the two hard spectra (B and C) directly, we fitted spectra A and D, and B and C with the same underlying model, and attempted to reproduce the different spectral shapes by applying additional absorption. 
Spectra A and D are well-fitted by a hot disc, with characteristic temperature of $\approx$1.8 keV. The very small normalisation of ($K_{\rm bb}\approx 340$) is indicative of small disc truncation radius. Spectrum D requires a multiplicative constant $K \approx$0.06, and additional uniform absorption of $(2.5\pm0.4) \times 10^{22}\, {\rm cm}^{-2}$. The observed fluxes in the 2-10 keV band from spectrum A and D are $\sim 3.6\times 10^{-8}\, {\rm erg\,cm^{-2}\,s^{-1}}$ and $2\times 10^{-9}\, {\rm erg\,cm^{-2}\,s^{-1}}$, respectively. Fitted individually, both spectra return best fit parameters consistent with those reported above, and neither is better described by a power law continuum.
We fitted spectrum B and C using a power law continuum, and we linked the power law photon index and normalisation across the two spectra, obtaining  a photon index of $\mathnormal{\Gamma} = 1.90 \pm 0.04$. Spectrum C required additional absorption of $(220^{+50}_{-30})\times10^{22}\, {\rm cm}^{-2}$, with a partial covering factor of $0.88\pm 0.02$. The observed fluxes in the 2--10 keV band from spectra B and C are $\sim 1.6\times 10^{-9}\, {\rm erg\,cm^{-2}\,s^{-1}}$ and $3\times 10^{-10}\, {\rm erg\,cm^{-2}\,s^{-1}}$, respectively.

We note that the above analysis is based on simple phenomenological models with the aim to provide clues regarding the nature of the accretion state(s) sampled by GRS~1915+105 after July 2018.

%---------------------
\begin{table*}
\centering
\caption{Best fit parameters for the spectra shown in Fig. \ref{fig:spectra}. The interstellar hydrogen column was fixed to $N^{\rm ISM}_{\rm H} = 5\times 10^{22}{\rm cm}^2$. From top to bottom in column 1: %interstellar absorption,
multiplicative constant $K$; local absorption $N^{\rm loc}_{\rm H}$; local partial covering fraction $PCF$; disc-blackbody temperature $T_{\rm bb}$; photon index $\mathnormal{\Gamma}$; observed flux $F$ in the 2--10 keV band. 
The parameters marked with $^{*}$ ($\mathnormal{\Gamma}$ in Spectra B and C, and $T_{\rm bb}$ in Spectra A and D, respectively), are linked, so that the same continuum is used to fit each pairs of spectra. The \textsc{xspec} models used have the form: $\textsc{tbfeo} \times \textsc{tb}\_\textsc{pcf}\times(\textsc{powerlaw})$ or $\textsc{tbfeo} \times \textsc{tb}\_\textsc{pcf}\times(\textsc{discbb})$.}
    \begin{tabular}{lcccc}
    \hline
    \textbf{Parameter}  &\textbf{A} & \textbf{B (Plateau 1)} & \textbf{C (Plateau 2)} & \textbf{D (soft phase)} \\ 
    \hline
    $K$                                &  1 (fixed)                &   1 (fixed)       &   1 (fixed)               &    $0.064\pm0.002$ \\
    %$N^{\rm ISM}_{\rm H}$ [$\times 10^{22}cm^{2}$]   &\multicolumn{4}{c}{5.0}                     \\
    $N^{\rm loc}_{\rm H}$ [$\times 10^{22}cm^{2}$]   &  -               &   -               &    $220^{+50}_{-30}$      &    $2.5\pm 0.4$  \\
    $PCF$                                     &  -                        &   -               &    $0.88\pm0.02$        &    1 (fixed)  \\
    $T_{\rm bb}$ [keV]                      & $1.94 \pm 0.04$ $^{*}$    &   -               &  -                        &   $1.94 \pm 0.03 ^{*}$     \\
    $\mathnormal{\Gamma}$                                &  -                        &  $1.91\pm0.03$ $^{*}$                         &  $1.91\pm0.03$ $^{*}$ &   -   \\
    $F$ [$\times 10^{-8}\,{\rm erg\, cm^{2}\,s^{-1}}$]        &  3.6              &     0.16                  &  0.03                 &  0.2      \\
    $\chi^2$/d.o.f.                         &  $134.03/119$             &     $158.45/159$  &   $12.89/12$              &  $165.69/138$            \\
    \hline 
    \end{tabular}
    \label{tab:para}
\end{table*}
%---------------------

\subsection{Radio--X-ray plane}

To better compare the current behaviour of GRS~1915+105 with its past behaviour, as well as with other BH transients, we placed it on the radio--X-ray plane. 
We measured the radio and X-ray fluxes corresponding to Plateau 1, Plateau 2, and the Soft Phase taking the mean in the three phases. 
Figure \ref{fig:RX_corr} shows the radio--X-ray points (in grey) for a number of BH transients considered in \cite{Motta2017a}, plus MAXI J1820+070 from \cite{Bright2020}. The error bars account for both the scatter in the values, and the uncertainty on the counts-to-flux conversion described above. 
GRS~1915+105 was traditionally considered an outlier on the radio--X-ray correlation that the vast majority of BH transients follow, as shown by the red dots in the figure, which mark the position that GRS~1915+105 occupied during its C state phases of its 27-years long outburst (points taken from \citealt{Rushton2010}). 
While during Plateau 2 and during the Soft Phase GRS~1915+105 still clearly lies away from the other sources on the plane, during Plateau 1 GRS 1015+105 falls on the correlation, incidentally approximately in the same position occupied by Cyg~X--1.

%%%%%%%%%%%%%%%%%%%%%%%%%%%%%%%%%%%%%%%%%%%%%%%%%%%%%%%%%%%%%%%%%%%%%%%%%%%%%%%%%%%%%%%%%%%%%%%%%%%%%%%%%%%%%%%%%%%%%%%%%%%%%%%%%%%%%%%
\section{Discussion}\label{sec:discussion}

We have presented the results of a comparative radio and X-ray study of GRS~1915+105, primarily focusing on the evolution of the source from MJD 58300 (July 2018) to MJD 59200 (November 2020). We are motivated by the fact that in 2018 the source underwent a flux decline in the X-rays that might have been interpreted as a transition to a canonical hard state, possibly preceding a long-expected quiescence (\citealt{Truss2006}).
Our data show that, despite the low X-ray fluxes displayed lately, GRS~1915+105 is still very active in radio, and during the last several months has been showing marked activity that is qualitatively very similar to that observed in the past. This provides evidence for two important facts: first, that the correlation between radio and X-ray emission that for many years characterised GRS~1915+105 (\citealt{Fender2004a}) ceased to exist sometime in June 2019; and second, that the X-ray behaviour alone--in the absence of such radio--X-ray correlation--is very misleading, as it offers only a partial view of the current state of the source. GRS~1915+105 has not entered quiescence, and might not be approaching it either (see also \citealt{Neilsen2020} and \citealt{Koljonen2021} for a discussion of this topic). Rather, the accretion processes that must be at work to feed the jets responsible for the observed radio behaviour are for some reason not directly visible to the observer. 

The low X-ray fluxes observed during Plateau 1, i.e. after the exponential decline occurred in 2018 (see also \citealt{Negoro2018}) appear as highly unusual for GRS~1915+105. Furthermore, the almost total lack of variability in the X-ray emission, as opposed to marked radio flaring observed after June 2019 that characterises Plateau 2 (referred to as the obscured phase by \citealt{Miller2020}) is certainly unprecedented. Based on the data reported here, and on older data published by other authors (e.g., \citealt{Klein-Wolt2002}) Plateau 2 is the longest, lowest flux, and hardest X-ray plateau ever observed in  GRS~1915+105, and the only one associated with marked radio flaring. Any previous plateau was a plateau both in radio and in the X-rays, and any Flaring Phase has occurred in both bands. 

Since the beginning of Plateau 2, absorption is a constant characteristic of the energy spectra of GRS~1915+105. Our spectral analysis results are fully consistent with those reported by previous works (\citealt{Koljonen2020,Miller2020,Neilsen2020,Balakrishnan2021,Koljonen2021}): an obscuring in-homogeneous medium is required to explain the average spectrum we extracted from Plateau 2. This agrees with the results from the Chandra spectra, which were taken when GRS~1915+105 was observed in a deep obscuration state \citep{Miller2020}. 
Absorption also impacted the occasional flares captured by \textit{NICER} during the obscured state\footnote{Some of these flares were also detected by BAT, when sufficiently long to be seen in the 1-day averaged light curve.}, which still reached a high luminosity, indicating that the intrinsic X-ray luminosity of GRS~1915+105 is likely not far from the Eddington limit  \citep{Neilsen2020}. The evolution of the X-ray emission during one particular flare reported by \cite{Neilsen2020}, shows that the flares are characterised by harder-when-brighter spectra, and require high density and variable in-homogeneous local absorption, properties very reminiscent of the behaviour of V404 Cyg during a vast majority of the flares observed in 2015 (\citealt{Motta2017b}). Also reminiscent of V404 Cyg are the spectral properties of the obscured state around the occasional flares observed. A high equivalent column density absorber, which was almost completely covering the emission from the central portion of the accretion flow, was responsible for the spectrally hard and low flux emission observed during a number of plateaus during the 2015 outbursts of V404 Cyg (\citealt{Motta2017a,Kajava2018}).

The behaviour and properties of GRS~1915+105 during the obscured state are consistent with those observed in all the systems displaying phases of strong, variable local absorption: all tend to show high-amplitude flares. Some noteworthy examples are V4641 Sgr \citep{Revnivtsev2002}, Swift J1858--0814 \citep{Hare2020,Munoz-Darias2020}, and even some Seyfert II Galaxies \citep{Moran2001}, but perhaps the best example remains that of V404 Cyg. 
In the case of V404 Cyg, the behaviour observed both during the low-flux phases and during flares has been ascribed to the presence of an inflated accretion disc. Such a thick disc was likely sustained by radiation pressure induced by high accretion rate episodes, and was fragmenting out in the clumpy outflow responsible for the local and variable Compton-thick absorption which affected the spectra. Something similar might be happening in GRS~1915+105 in the obscured state, even though the obscured state in this case is significantly more extended than for V404 Cyg. 

Owing to the better energy resolution and signal-to-noise ratio achieved with Chandra and \textit{NICER}, respectively, observations of GRS~1915+105 in the obscured state offered a deeper insight into the properties of this state. Both \citet{Neilsen2020} and \citet{Miller2020} showed that the absorber in GRS~1915+105, besides being in-homogeneous, requires a multi-temperature profile, and it is likely radially stratified. On the one hand, \citet{Miller2020} showed that at least two types of media form this absorber: an inner bound (failed) magnetic wind, and an outer, cooler component which could be a thermally-driven outflow. In this scenario, the failed wind would be responsible for the obscuration. On the other hand, \cite{Neilsen2020}, who analysed data taken at a different time, showed that their results are consistent with the presence of radially stratified puffed-up outer disc, which would hide the regions closer to the BH. It seems therefore plausible that the absorber in GRS~1915+105 covers a large portion of the accretion disc, from a few hundreds gravitational radii, to the outer disc, at radii larger than tens of thousands $R_{\rm g}$. In this respect GRS~1915+105 differs from V404 Cyg. In the latter both neutral and ionised winds were being launched from the outer disc, but a cold, optically thick and partially covering absorbing material was launched from within a few hundreds of gravitational radii from the black hole, and was clearly outflowing with a velocity of the order 0.05~c \citep{Motta2017b}.

Considering GRS~1915+105 in the context of the radio--X-ray plane, we see that for the first time since its discovery, during Plateau 1 the system was consistently located on the correlation traced by the majority of known BH transients. During both Plateau 2 and the Soft Phase the position of GRS~1915+105 in the radio--X-ray plane confirms that a large fraction of the X-ray flux is lost. If, for the sake of the argument, we assume that the correlation holds during Plateau 2, we estimate that the obscuration of the inner accretion flow might be causing a loss of approximately a factor $\sim$200 in X-ray luminosity, consistent with the obscuration scenario. However, note that the radio--X-ray correlation strictly holds only during the canonical hard state (\citealt{Fender2004a}), when a compact steady jet is detected, and the radio and X-ray emission can be directly compared. The above estimate is intended only as an indication of the overall behaviour of the source, the emission being so heavily modified by absorption, we have no means to confirm the X-ray state of the source.

Based on our results, it seems that Plateau 1 was a truly dim state, very similar to the canonical low-flux hard state typical of more well-behaved BH transients, characterised by low X-ray and radio flux, as well as low long-term variability in both bands. Such a state is certainly unusual for GRS~1915+105, which has never been observed before in a low-luminosity canonical hard state \citep[but see][]{Gallo2003}. Instead, despite the fact that Plateau 2 has been dubbed `the obscured state', it is not really a state, but rather a \textit{condition} directly dependent on the presence of local absorption along our the line of sight. 
The accretion processes that must be feeding the markedly variable jet observed in radio must be happening beneath a complex layer of material local to the source, which shields the inner part of the accretion flow, and thus blocks a large portion of the X-ray emission. Behind this Compton-thick curtain of material, GRS~1915+105 is most likely evolving through various states and transitions, consistent with what it had been doing for 25 years until June 2018. 
This means that perhaps, as already proposed by \cite{Miller2020}, GRS~1915+105 did not really enter the outburst phase in 1992, when it was discovered, but simply emerged from an obscured phase similar to the one we are witnessing at the time of the observations presented here.

GRS~1915+105 is an important system for a number of reasons, including being in many ways a small-scale AGN. As in the case of many other Galactic BHs, its different states may correspond to a number of AGN classes, but the specific multi-band behaviour of GRS~1915+105 has clear counterparts in AGN \citep{Miller2020}. %\MG{$\leftarrow$ not really sure about this sentence}. 
So, how does this new observed state of GRS~1915+105--heavily absorbed in X-rays and active in radio--compare to AGN?
In AGN, different absorbers on different scales affect the X-ray spectrum: from dust lanes in the host galaxy, to the parsec-scale torus, to accretion-disc scale clouds/winds \citep[see][for a recent review]{2017NatAs...1..679R}. 
Large column densities $N_{\rm H}\sim 4{-}9\times 10^{23}$ cm$^{-2}$, comparable to those measured in the X-ray spectra of GRS~1915+105 in Plateau 2, are inferred from hard X-ray observations of local radio galaxies \citep[see][and references therein]{2018MNRAS.474.5684U} and young radio sources \citep[e.g., Mrk 668,][]{2004A&A...421..461G}.
The radio jet activity of young radio sources is also intermittent, possibly due to the interaction with the X-ray absorbing dense circumnuclear medium. However, most of the X-ray absorbers found in these sources are likely found on parsec scales and beyond. 
Compton-thick absorbers on accretion disc--scales are observed in some Changing-Look AGN \citep[e.g.,][]{2005ApJ...623L..93R}, but their time scales for variability, once scaled down to stellar mass black holes, are much shorter than what is observed in GRS~1915+105. Thus, a phenomenon similar to the one observed in GRS~1915+105 has not been identified in AGN, yet.

Finally, while accreting super-massive BHs and stellar-mass black holes are connected by the same fundamental physics \citep{Merloni2003, Falcke2004}, which governs their inner accretion flow, their larger-scale structure differ quite appreciably. In particular, accreting stellar mass BH have a companion star, the behaviour of which can potentially greatly affect the long-term behaviour of the accretion disc. \cite{Neilsen2020} speculated that if a vertically extended outer accretion disc is responsible for the obscuration in GRS~1915+105, its formation might have been triggered by changes in the companion star, e.g. an increase in the mass supply into the accretion disc, which necessarily would trigger a change in the accretion flow from the outside-in. The fact that an equivalent of the obscured state seen in GRS~1915+105 has not been seen in AGN might support this hypothesis. 

%%%%%%%%%%%%%%%%%%%%%%%%%%%%%%%%%%%%%%%%%%%%%%%%%%%%%%%%%%%%%%%%%%%%%%%%%%%%%%%%%%%%%%%%%%%%%%%%%%%%%%%%%%%%%%%%%%%%%%%%%%%%%%%%%%%%%%%
\section{Summary and conclusions}\label{sec:conclusions}

In 2018, the black hole binary GRS~1915+105, after 25 years of high-luminosity X-ray activity, decayed to a prolonged low-flux X-ray state. Due to this relatively sudden change in the X-ray behaviour of GRS 1915+105, some were led to believe that its outburst, the longest ever observed from a black hole X-ray binary, was approaching its end.

We analysed the simultaneous X-ray and radio data collected over almost 3 decades with various facilities, focusing on the most recent evolution of the system. Our data show that at the beginning of the dim X-ray state GRS~1915+105 was also relatively faint in radio. Since June 2019 the system has been showing marked radio activity, characterised by the signatures of relativistic jets, and X-ray spectra affected by high and variable in-homogeneous absorption. Our results show that more recently GRS~1915+105, while still affected by heavy absorption, transitioned to a softer state, which was accompanied by a decrease in the radio flaring that resumed when GRS~1915+105 moved back to a hard(er) state. 

We argue that GRS~1915+105 first transitioned to a low-luminosity hard state, similar to the canonical hard state shown by many other black hole X-ray binaries, and then entered a prolonged obscured phase. In this phase the highly variable radio jets we have been observing for months must be fed by the same sort of accretion processes that have been seen often in the past in GRS~1915+105, and are now happening behind a complex layer of absorbing material. 
We therefore conclude that GRS~1915+105 is far from being in quiescence, even though a substantial change in the accretion flow--perhaps the launch of a powerful outflow and/or the thickening of the outer disc--must have occurred at some point around June 2019. 
The behaviour of GRS~1915+105 in the obscured state appears to have no counterpart in its super-massive relatives, the AGN, where the time-scales typical of similar radio-bright obscured phases (once they are scaled up in mass) are either much longer, or much shorter than in GRS~1915+105. 

%%%%%%%%%%%%%%%%%%%%%%%%%%%%%%%%%%%%%%%%%%%%%%%%%%%%%%%%%%%%%%%%%%%%%%%%%%%%%%%%%%%%%%%%%%%%%%%%%%%%%%%%%%%%%%%%%%%%%%%%%%%%%%%%%%%%%%%
\section*{Acknowledgements}

%--- Financial suppeort
SEM acknowledges the Violette and Samuel Glasstone Research Fellowship programme, and the UK Science and Technology Facilities Council (STFC) for financial support. SEM and DW acknowledge the Oxford Centre for Astrophysical Surveys, which is funded through generous support from the Hintze Family Charitable Foundation.
JJEK acknowledges support from the Academy of Finland grant 333112 and the Spanish MINECO grant ESP2017-86582-C4-1-R.
MG is supported by the ``Programa de Atracci\'on de Talento'' of the Comunidad de Madrid, grant number 2018-T1/TIC-11733.
MDS and AS acknowledge financial contribution from the agreement ASI-INAF n.2017-14-H.0 and from the INAF mainstream grant. PAW acknowledges financial support from the University of Cape Town and the National Research Foundation. We also acknowledge support from the European Research Council under grant ERC-2012-StG-307215 LODESTONE.\\
%---Observatories
We thank the staff of the Mullard Radio Astronomy Observatory, University of Cambridge, for their support in the maintenance and operation of AMI. 
We thank the staff at the South African Radio Astronomy Observatory (SARAO) for scheduling the MeerKAT observations. The MeerKAT telescope is operated by the South African Radio Astronomy Observatory, which is a facility of the National Research Foundation, an agency of the Department of Science and Innovation. 
This research has made use of MAXI data provided by RIKEN, JAXA and the MAXI team \citep{Matsuoka2009}. \\
%--- Others
All the authors wish to heartily thank Guy Pooley, who sadly recently passed away. His work was instrumental for the understanding of the radio properties of GRS~1915+105.
%

%%%%%%%%%%%%%%%%%%%%%%%%%%%%%%%%%%%%%%%%%%%%%%%%%%%%%%%%%%%%%%%%%%%%%%%%%%%%%%%%%%%%%%%%%%%%%%%%%%%%%%%%%%%%%%%%%%%%%%%%%%%%%%%%%%%%%%%
\section*{Data Availability}

The un-calibrated MeerKAT visibility data presented in this paper are publicly available in the archive of the South African Radio Astronomy Observatory at https://archive.sarao.ac.za, subject to a standard proprietary period of one year. 
Data from the \textit{Swift}/BAT and the RXTE/ASM are publicly available in the NASA's HEASARC Data archive. The MAXI/GSC data are made available by RIKEN, JAXA and the MAXI team. Data that are not available thorough public archives, and all source code, will be shared on reasonable request to the corresponding author.



\bibliographystyle{mnras.bst}
\bibliography{biblio}

%\begin{thebibliography}{99}
%\end{thebibliography}

%%%%%%%%%%%%%%%%%%%%%%%%%%%%%%%%%%%%%%%%%%%%%%%%%%

%%%%%%%%%%%%%%%%% APPENDICES %%%%%%%%%%%%%%%%%%%%%

%\appendix



%%%%%%%%%%%%%%%%%%%%%%%%%%%%%%%%%%%%%%%%%%%%%%%%%%

% Don't change these lines
\bsp	% typesetting comment
\label{lastpage}
\end{document}